# OPTIMIZATION OF DECENTRALIZED SCHEDULING FOR PHYSIC APPLICATIONS IN GRID ENVIRONMENTS

**Florin Pop***

* Faculty of Automatics and Computer Science, University "*Politehnica*" of Bucharest
E-mail: **florinpop@cs.pub.ro**

***Abstract.*** This paper presents a scheduling framework that is configured for, and used in physic systems. Our work addresses the problem of scheduling various computationally intensive and data intensive applications that are required for extracting information from satellite images. The proposed solution allows mapping of image processing applications onto available resources. The scheduling is done at the level of groups of concurrent applications. It demonstrates a very good behavior for scheduling and executing groups of applications, while also achieving a near-optimal utilization of the resources.

## 1. INTRODUCTION

Over the last years the processing power of the computing machines has grown enormously, due to the massive development of the semiconductor industry. Tasks that on older computers were virtually impossible to achieve in a reasonable time, could now be easily solved on a medium–to–high end personal computer.

One of the most important development was the invention of high-speed computer networks. Local area networks (LANs) allow hundreds of machines within a building to be connected in such a way that small amounts of information can be transferred between machines in a few microseconds or so (rate is 10 to 1000 million bits/sec) [1]. The result of technologies is that it is now easy to put together computing systems composed of large numbers of computers connected by a high-speed networks. They are usually called computer networks or distributed systems.

Distributed systems have become very useful especially in the case of scientific applications, where there is necessary the processing of a very large data volume, in a very short amount of time, as well as the storage of these data.

However, the computer capabilities have been greatly expanded with the development of the networking technologies and the Internet. Nowadays a large amount of software is being developed to take advantage of all the communication features of the Internet: Client - Server services, Information exchange, Multimedia streaming, Concurrent computation, and the list can be expanded. Among all these services, this section will focus on the concurrent computation technologies, which best match with this project area of interests. Concurrent computation (or distributed computing) is defined by a complex processing that can be split into simpler ones, each being solvable by a computer, and the partial results could be merged into the larger solution of the initial processing. Concurrent computation over the Internet involves developing software for the side responsible with the division of the problem into multiple ones, passing the packets of data to the "worker" side, and then collecting and combining the results; the other side represents the

## 2. DEDICATED APPLICATIONS FOR GRID SYSTEMS

I start with a questions: What type of computational problems is solvable by application of newly developing infrastructure? And the answer is: The simplest situation is large amount of small, mutually independent jobs. Typical example can be parametric studies in which the same calculation is repeated for many times with slightly modified parameters. Results of such calculations are combined and used for evaluation of impact of parameter change (e.g. rigidity of a used glue) for the properties of the whole system. Grid is the optimal environment for such types of calculations. A user prepares large batch – that can contain even million of independent jobs – and grid environment will take care of their scheduling, running, results collecting and potentially also final processing. Grid will also handle the crashed jobs that can be automatically submitted again.

Similar type of jobs is treatment of extensive data files "piece after piece". Simple example is digital smoothing of satellite images. Each image can be treated individually, it is possible to split large images into smaller ones, individual parts can be handled separately and the result can be merged into a single image again. This (parallel) processing is the most interesting type of jobs for which Grids have been developed.

These types of jobs are extensive jobs whose processing using one CPU would take incredibly long time in general especially if their memory requirements are too big. Classical approach for solving these jobs is to use a special – and very expensive – parallel computers. The job is simultaneously processed by large number of CPUs that have access to very large memory (if we have one thousand of CPUs each with one GB memory, we have together considerable 1 TB of internal memory accessible for one job).

Parallel computers are basically nothing more than a set of processors and memories interconnected by very fast network especially with special properties. A computer cluster (group of computers interconnected by "standard", still very fast, network and located in a small space) is the same at a specific level of abstraction and on a higher abstraction level we talk about a Grid (distance among the computers is higher in several orders of magnitude compared to a cluster).

Programming for parallel computers is unfortunately very hard because the job has to be split in a such a way that each processor has to be busy and the processors do not wait for each other one. The speed of data transfer inside a parallel computer is extremely high. This enables exchange of large volumes of data among processors without the loss of total performance. The situation in grid environment is much more complicated. The distance among processors and used network - high-speed Internet - cause much higher price (means time) of data transfer. Algorithms and procedures that are used in parallel computers are not sufficiently effective in grid environment. Therefore they have to be substituted by new ones or some of jobs types can not be currently effectively solved using Grid.

The basic term concerning parallelization of a job is the granularity of the problem. If the problem is highly dependent on the result of other sub-problems, with fine-grained parallel calculations. This is the case, for example, in a calculation of the weather, which can be split into many smaller calculations of the weather in small volumes of the atmosphere. Each of these calculations is strongly affected by what is happening in neighboring volumes. In fact, even changes in the weather very far away can have an impact. In practice, the transformation of such job types for grid environment is highly difficult and fine-grained parallel calculations require very clever programming to make the most of their parallelism, so that the right information is available to processors at the right time.

At the opposite end of the granularity scale are coarse-grained or embarrassingly parallel calculations, where each sub-problem is independent of all others. This is the case, for example for so-called Monte Carlo simulations, where you vary the parameters in a complex model of a real system and study the results using statistical techniques - a sort of computer experiment used for example in computational chemistry. Each calculation can be done independently of the others in this case. Also the mentioned example of digital images processing belongs to the same category. As a rule of thumb, fine-grained calculations are better suited to big, monolithic parallel supercomputers, or at least very tightly coupled clusters of computers, which have lots of identical processors with an extremely fast, reliable network between the processors, to ensure there are no bottlenecks in communications.

This type of computing is often referred to as high-performance computing. On the other hand, embarrassingly parallel calculations are ideal for a more loosely-coupled network of computers, since delays in getting results from one processors will not affect the work of the others. These types of calculations are often referred to as high-throughput computing. Naturally, the grid environment prefers coarse-grained parallel calculations. But in fact, many of the interesting problems in science require a combination of fine- and coarse-grained approaches as homogeneous coarse-grained algorithms are very rare. And this is where the Grid can be particularly powerful due to availability of suitable computational resources for fine-grained sub-jobs whose results can be combined by the manner of coarse-grained sub-jobs. Real example of described approach is doing complex climate modeling of the Earth, when researchers want to see how the calculations depend on different parameters in their models. So they want to launch many similar calculations. Each one is a fine-grained parallel calculation, because predicting climate is like predicting the weather on a longer time scale. So each calculation needs to be run on a single cluster or supercomputer. However, the many independent calculations could be distributed over many different clusters on the Grid, thus adding coarse-grained parallelism and saving a lot of time.

Suitable decomposition of a computational job followed by assignment of the sub-jobs to corresponding parts of the Grid is the area currently changing from an art - controlled by small group of masters – through (a lot of human work) to engineering discipline whose tools are available for growing group of users. It is expected that the major part of the forming infrastructure will be exploited, especially in the first phase of the project, by applications from physics of elementary particles, HEP (High Energy Physics), that is in the Czech republic represented by the Institute of Physics of Czech Academy of Sciences. For HEP community is Grid necessary and in essence the only solution how to store and process data in PB ($10^{12}$) orders of magnitude from particle physics experiments (Atlas, CMS, Alice, LHCb and other). Concurrently Grid enables cooperation in really worldwide scale - after all worldwide web has its roots also in CERN, the centre of HEP research.

One of the key aim of the EGEE-II (8Enabling Grids for E-sciencE) project is at the same time broadening of the users foundation through wide spectrum of potential applications. The portfolio of problems solvable using grid environment is relatively broad but their usage in grid environment is usually hindered by insufficient readiness on both sides - Grid specialists and users themselves. Taking into account described types of computational problems, the characteristic applications include computational chemistry simulations of biologically significant systems, processing of bioinformatics and medical data, material simulations and tests, growing interest is from monitoring and processing Earth observation data, distribution and analysis of astronomical measurements and particle physics experiments, and especially recently increased interest from real-time video and image processing [4].

## 3. OPTIMIZATION SOLUTION FOR TASK SCHEDULING IN GRID SYSTEMS

In this approach, the problem of task balancing is more important for an optimal scheduling. In dynamic environment can appear changes in time. A distributed scheduler can be able to resolve the fault tolerance problem and to be adaptable in the existing conditions. The optimization solution that will be proposed will develop a scheduler that could be able to run with a minimal grid requirements. The scheduling algorithm will use genetics algorithms. The scheduling system will work with groups of tasks in a distributed system using the existing formwork for centralized scheduling an monitoring. In this research, the computers in the grid are part of a group, according to their specific function, as described below [2]:

**Scheduling Group**: The computers in this group run the scheduling algorithm. They receive requests for tasks to be scheduled, and return the best planning according to the genetic algorithm.
**Execution Group**: The computers in this group execute the tasks that had been previously scheduled and assigned to them.

One major objective of this research is to fit in real scenarios of grid computing. I will extend my application to a general level in which two scenarios can be perfectly used in task scheduling applications. In the first scenario the batches of tasks to be allocated are received from remote sites from users by means of a portal. The user requests for task allocation are sent to the computers that run the scheduling algorithm.

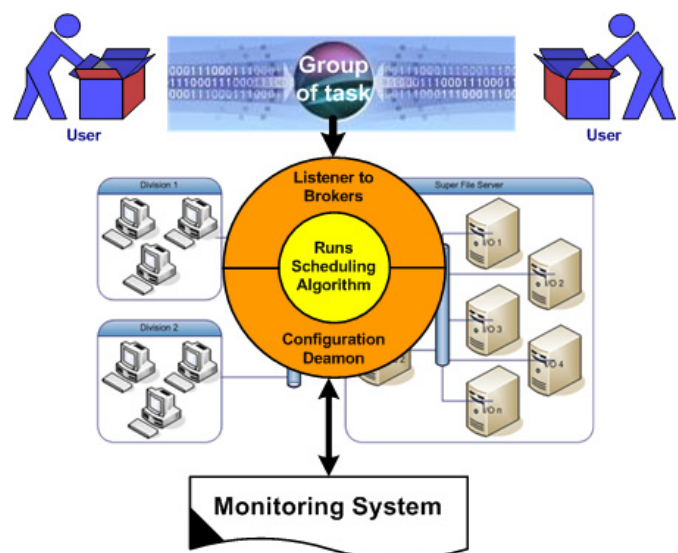

*Figure 1. System Architecture*

There are therefore some computers designated to receive user requests (brokers), and some different entities that run the scheduling algorithm and find the best schedule (agents). In this case, the brokers provide the input for this scheduling algorithm, but do not actually run it.

In distributed processing systems such as the Grid, the fundamental problem of the maximization of system performance through task mapping must be based on a reliable and efficient scheduling algorithm. A good allocation algorithm may improve resource utilization and increase significantly the throughput of the system. Furthermore, because I've tried to implement a practical scheduler which makes it's decisions taking into consideration this proposed algorithm needed to use the received parameters about the state of the distributed algorithm with the purpose to produce the desired result with minimum expenditure and also to react on new external and internal conditions [3].

In this particular problem, using a genetic algorithm implies finding a near-optimal solution which is considered satisfactory, based on the predefined performance evaluators. In the initialization stage, in order to have a better probability to find a near-optimal solution, various start-points are generated. This is achieved by using different probabilistic distributions which are used by each agent for population initialization (Poisson, Normal, Geometrical, Uniform, Laplace).

### 3. PRELIMINARY RESULTS

In order to test scheduling system, an experimental cluster with 11 computing resources was configured. The genetic algorithm was tested using groups of tasks ranging from 50 to 100 simultaneous tasks. The experiment had the purpose to compare the results obtained by the current strategy used in real-time scheduling, First Come First Served (FCFS), and the Genetic Algorithm (GA) carried out with a number of three agents. The tests were performed over 200 generations, on a set of 100 incoming tasks. The figure shows better results for the second method, when lower execution times on processors were obtained. The minimization of the execution times is a consequence of the approximately uniform load on processors achieved by the Genetic Algorithm, as further detailed.

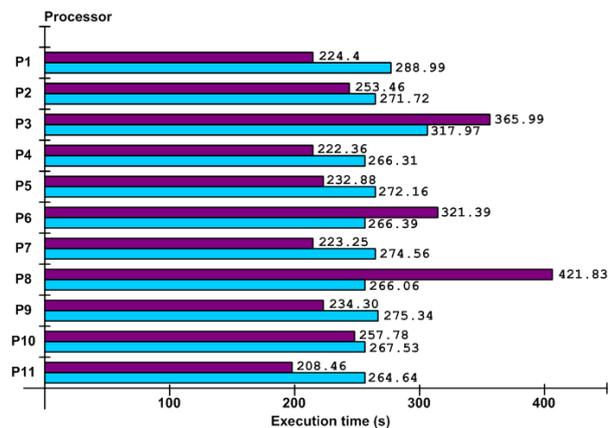

*Figure 2. Execution times with First Come First Serve method and Genetic Algorithm method*

### 4. CONCLUSIONS AND FUTURE WORK

This paper describes a genetic scheduling approach, referring to a scalable, decentralized agent-based strategy for the problem of task allocation in a dynamic, distributed environment. The experimental results obtained demonstrate that the optimization of the scheduling process in physics application projects can be achieved by considering two main criteria, as previously described. In addition, the decentralized strategy represents a scalable and reliable method that can be used in real-time distributed systems, such as the satellite image processing Grids. I also take into consideration various requirements that inevitably appear in a real scenario: deadline restrictions and resource limitations.